\documentclass[aps,twocolumn,pre,superscriptaddress,preprintnumbers,amsmath,amssymb]{revtex4-2}
\usepackage{graphicx}
\usepackage{color}

\begin{document}

\title{Tetratic Phase in 2D Crystals of Squares}

\author{Robert L\"offler}
\affiliation{Department of Physics, University of Konstanz, 78467 Konstanz, Germany}

\author{Lukas Siedentop}
\affiliation{Department of Physics, University of Konstanz, 78467 Konstanz, Germany}

\author{Peter Keim}
\email{peter.keim@uni-duesseldorf.de}
\affiliation{Department of Physics, Heinrich-Heine-Universit\"at D\"usseldorf, 40225 D\"usseldorf, Germany}
\affiliation{Institute for the Dynamics of Complex Systems, University of G\"ottingen, 37077 G\"ottingen, Germany}
\affiliation{Max-Planck-Institute for Dynamics and Self-Organization, 37077 G\"ottingen, Germany}

\date{\today}

\begin{abstract}
Melting in 2D is described by the celebrated Kosterlitz-Thouless-Halperin-Nelson-Young (KTHNY) theory. The unbinding of two different types of topological defects destroys translational and orientational order at different temperatures. The intermediate phase is called hexatic and has been measured in 2D colloidal monolayers of isotropic particles. The hexatic is a fluid with six-fold quasi-long-ranged orientational order. Here, the melting of a quadratic, 4-fold crystal is investigated, consisting of squares of about $4 \times 4\;\mu\mathrm{m}$. The anisotropic particles are manufactured from a photoresist using a 3D nanoprinter. In aqueous solution, particles sediment by gravity to a thin cover slide where they form a monolayer. The curvature of the cover slide can be adjusted from convex to concave, which allows to vary the area density of the monolayer in the field of view. For low densities, the squares are free to diffuse and form a 2D fluid while for high densities they form a quadratic crystal. Using a four-fold bond-order correlation function, we resolve the tetratic phase with quasi long ranged orientational order.
\end{abstract}

\maketitle

\section{Introduction}

Melting in 2D is described by the celebrated Kosterlitz-Thouless-Halperin-Nelson-Young (KTHNY) theory \cite{Kosterlitz1972,Kosterlitz1973,Nelson1977,Halperin1978,Nelson1979,Young1979}. Starting with an idealized mono-crystal, topological defects like dislocations can be created in pairs with antiparallel Burgers vector. The dissociation and diffusion of unbound dislocations marks the point, where shear elasticity disappears and the ensemble enters a fluid phase. On the crystal side, a six-fold director field can be associated to each particle given by the positions of the nearest neighbours. For isotropic particles, this is given by the fact that the closed packed crystal structure is hexagonal, thus each particle has six nearest neighbours. Dislocations are most easily detected by Voronoi tessellation, marking a pair of particles having five and seven nearest neighbours. Those arrangements stem from thermal excitation and a local displacement of the lattice. Since dislocations can be excited everywhere within the 2D bulk (not only at the surface/border of the crystal) the transition is predicted to be continuous.\\

As pointed out by Nelson and Halperin, this fluid phase is not yet isotropic on temporal averaging the positions of the particles under Brownian motion. Dislocations effectively destroy the translational order but not the orientational one: the local director field can be correlated as function of distance and one finds a rather slow decay being best fitted with an algebraic function. This algebraic decay of orientational order is named quasi-long ranged and together with the short ranged translational order, the thermodynamic phase of this anisotropic fluid is called hexatic. A second class of topological defects, namely disclinations have to appear to destroy orientational order completely. Based on Voronoi tessellation, this is when dislocations dissociate into isolated fivefold and isolated sevenfold coordinated particles at higher temperatures (or lower 2D-density, respectively). Those disclinations cause the bond-order correlation to decay very fast, best fitted with an exponential function and the very fast decay of the orientational order marks the isotropic fluid.\\

The two step melting in 2D with an intermediate hexatic phase was tested in simulations \cite{Jaster1998,Mak2006,Lin2006} and experiment. The latter almost exclusively with colloidal ensembles, where the particles can be monitored in time with video microscopy \cite{Murray1987,Tang1989}. Long ranged dipolar particle interactions are completely in line with theoretical predictions \cite{Kusner1994,Kusner1995,Marcus1996,Zahn1999,Zahn2000,Eisenmann2005,Zanghellini2005,Keim2007,Gasser2010} while short ranged interactions, e.g. for hard discs show a weakly first order hexatic to isotropic transition \cite{Thorneywork2017,Thorneywork2018}, in line with computer simulations \cite{Bernard2011,Engel2013,Kapfer2015}. Besides this, the picture does not change qualitatively under small anisotropy of the crystal given by an outer in-plane field \cite{Eisenmann2004,Eisenmann2004b,Froltsov2005}.\\

The question naturally arises what happens with the KTHNY-scenario if not hexagonal crystals are molten: interestingly in his awarded work, Mike Kosterlitz started the renormalization group analysis with a square lattice \cite{Kosterlitz1973}. Will an analogon of the hexatic phase, namely a tetratic phase as thermodynamic fluid phase with four-fold quasi-long range order show up upon melting? Square lattices are most easily made of squares and in computer simulations of rather small systems consisting of $196$ and $784$ particles, a tetatric phase was indeed observed \cite{Wojciechowski2004}. The results were not completely conclusive about the first or second order characteristic of the tetratic–isotropic transition. They did also not completely rule out a columnar phase, where strict periodicity is only given in one direction but perpendicular to this, displacements of fractions of lattice spacing between columns are allowed. This phase is not easily distinguished from defect-rich cubic crystals known in 3D, where vacancies are allowed to dislocate about several lattice spacing in one direction due to the shape of the cubes \cite{Smallenburg2012}. An experiment with about $500$ macroscopic particles (roughly $8 \times 8~\mathrm{mm}^2$) on a vibrating plate again indicated a tetratic phase but no columnar phase. A colloidal ensemble of $2 \times 2~\mathrm{\mu m}^2$ squares fabricated of an UV-illuminated photoresist with a structured mask showed a completely different scenario: the author report a rotator crystal, where the squares are occupying the sites of a hexagonal crystal but orientation of the square (body-orientation) is random in space and time \cite{Zhao2011}. Even more surprising, at larger packing fractions they report a rhombic crystal out of squares. Here, the symmetry of the body is not reflected in the symmetry of the crystal. However, shortly later this puzzle was resolved: Using the concept of hypercubes with rounded edges it was shown in simulations that rhombic symmetry and rotator crystals are functions of shape when a perfect square is degenerated to a circular disk \cite{Avendano2012}. Tetratic phases can also be build of domino shaped particles with aspect ratio $1:2$ where two rectangles form a square \cite{Donev2006} or wider aspect ratios \cite{Dertli2024}.\\

\section{Fabrication of squares}

Care has to be taken to manufacture squares with sharp enough edges if one is experimentally interested in the tetratic phase. In this work, we printed squares of $4 \times 4~\mathrm{\mu m}^2$ with a thickness of $2~\mathrm{\mu m}$ using direct laser writing in a photoresist with a commercial setup (nanoscribe GT). An infrared femtosecond laser is focused with an $100~\times$ objective into an ultraviolet photoresist (IP-Dip from nanoscribe). This way, only two-photon absorption develops the resist and structures smaller than the wavelength can be written. We used the dip in technique, where the microscope objective directly dips into the resist and squares are written onto a $30\times30~\mathrm{mm}^2\times700~\mathrm{\mu m}$ fused silica substrate. The laser focus has some overlap with the glass, such that particles stick to the surface of the cover slide. After writing the squares on the substrate, the non-illuminated resist is removed with developer ($10~\textrm{min}$ in MR-Dev 600 from nanoscribe and $5~\textrm{min}$ in isopropyl alcohol).\\%

\begin{figure}[h]
\centering
  \includegraphics[width=0.5\textwidth]{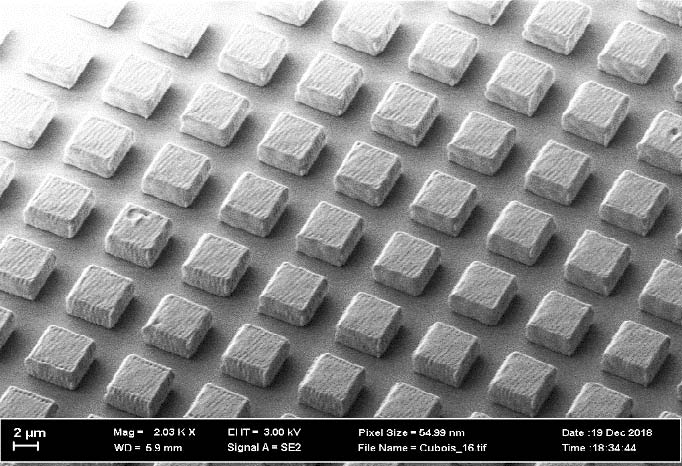}
  \includegraphics[width=0.5\textwidth]{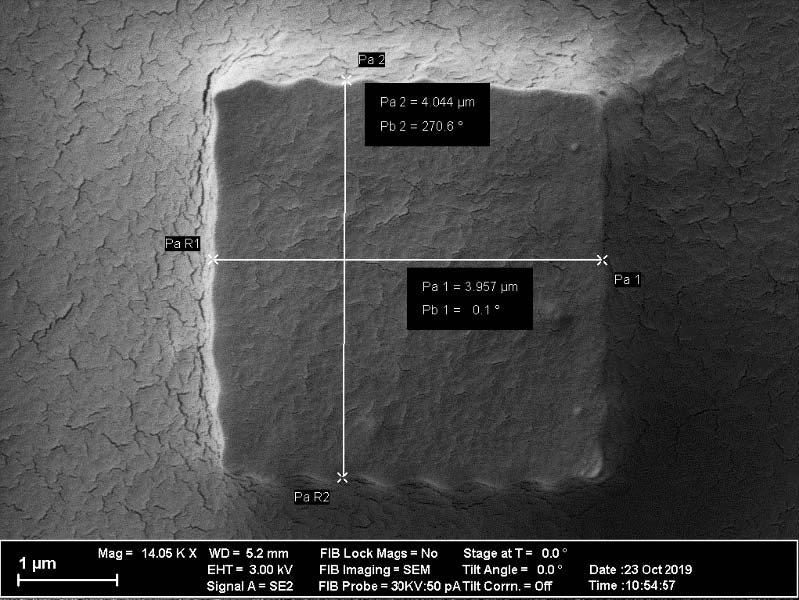}
  \caption{Electron micrographs of squares written on a substrate after developing. The upper image serves as an overview to show the arrangement of the cubes after printing, and the lower images shows a single cube to measure its size, which is $4\pm~0.04 \times 4\pm~0.04~\mathrm{\mu m}^2$.}
  \label{fig:EM_squares}
\end{figure}

Figure \ref{fig:EM_squares} shows electron microscopy images of the squares on the substrate. The surface of the squares shows a wavy structure (like a cracker) which is due to the printing process, where the laser focus is scanned across the sample in equidistant parallel lines. The path of the laser focus can be adjusted to make the surface more even, which comes at the expense of writing time. With the given laser intensity, beam velocity, and path we were able to write about four million squares within four days. The thickness of the squares of $2~\mathrm{\mu m}$ is the best compromise for particles i) being heavy enough in aqueous solution to sediment to form a monolayer, ii) thin enough to not stand vertically on the $2 \times 4~\mathrm{\mu m}^2$ side, and iii) thick enough that the probability for an overlap on top of each other is negligible. To remove the particles from the cover slide, a solution of
$1~\%_{wt}$ concentration of Pluronics-F127 and $0.3~M\;NaOH$ is used to wet the particles and cover slide for about $24~\mathrm{h}$. $NaOH$ will partially saponificate the PMMA based photoresist and Pluronics-F127 introduces sterical stabilization and helps wetting particles and cover slide for lift off. Taking a $1~\mathrm{ml}$ pipette of $1~\%_{wt}$ Pluronics-F127 in distilled water, spreading and sucking in the solution repeatedly will catch a finite fraction of squares from the glass plate into the solution. Putting the solution in a $1~\mathrm{ml}$ vessel, sonification for $10~\textrm{min}$, centrifuge $400~\mathrm{g}$ for $15~\textrm{min}$, removing supernatant containing $NaOH$ and refilling with $1~\%_{wt}$ Pluronics-F127 solution washes away the base, leading to aggregation of the colloids. After repeating about three times, $20\%$ to $30\%$ of the particles will be left over in about $1~\mathrm{ml}$ solution. This colloidal solution is used to fill the sample cell for further processing.\\

\section{Formation of a 2D mono-layer}

\begin{figure}[b]
\centering
  \includegraphics[width=0.2\textwidth]{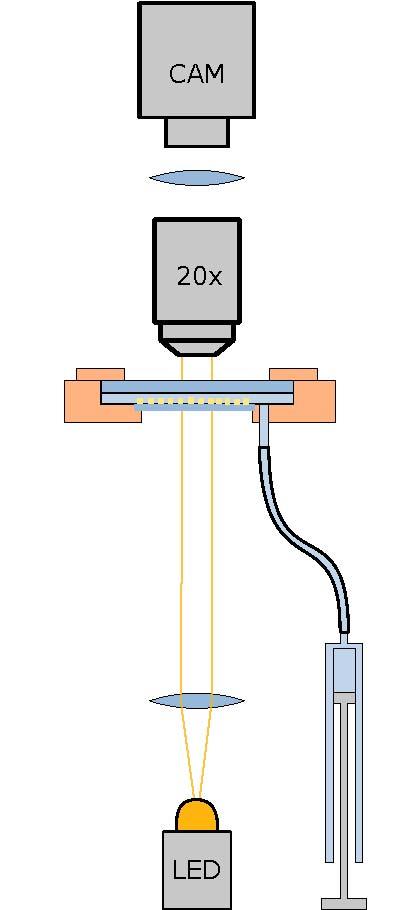}
  \caption{Sketch of the setup. The colloidal monolayer (yellow squares, not to scale) sediments within a layer of $2~\mathrm{mm}$ aqueous solution is sandwiched between two glass palates: The upper one is $2~\mathrm{mm}$ thick and will not bend, while the curvature of the lower glass plate (thickness of $80~\mu\mathrm{m}$) can be adjusted easily with an amplitude up to $100~\mu\mathrm{m}$ in the middle of the plate being $20~\mathrm{mm}$ in diameter. This allows to change the volume of the aqueous solution by a micro-stage driven syringe.}
  \label{fig:setup}
\end{figure}
\begin{figure}[t]
\centering
  \includegraphics[width=0.4\textwidth]{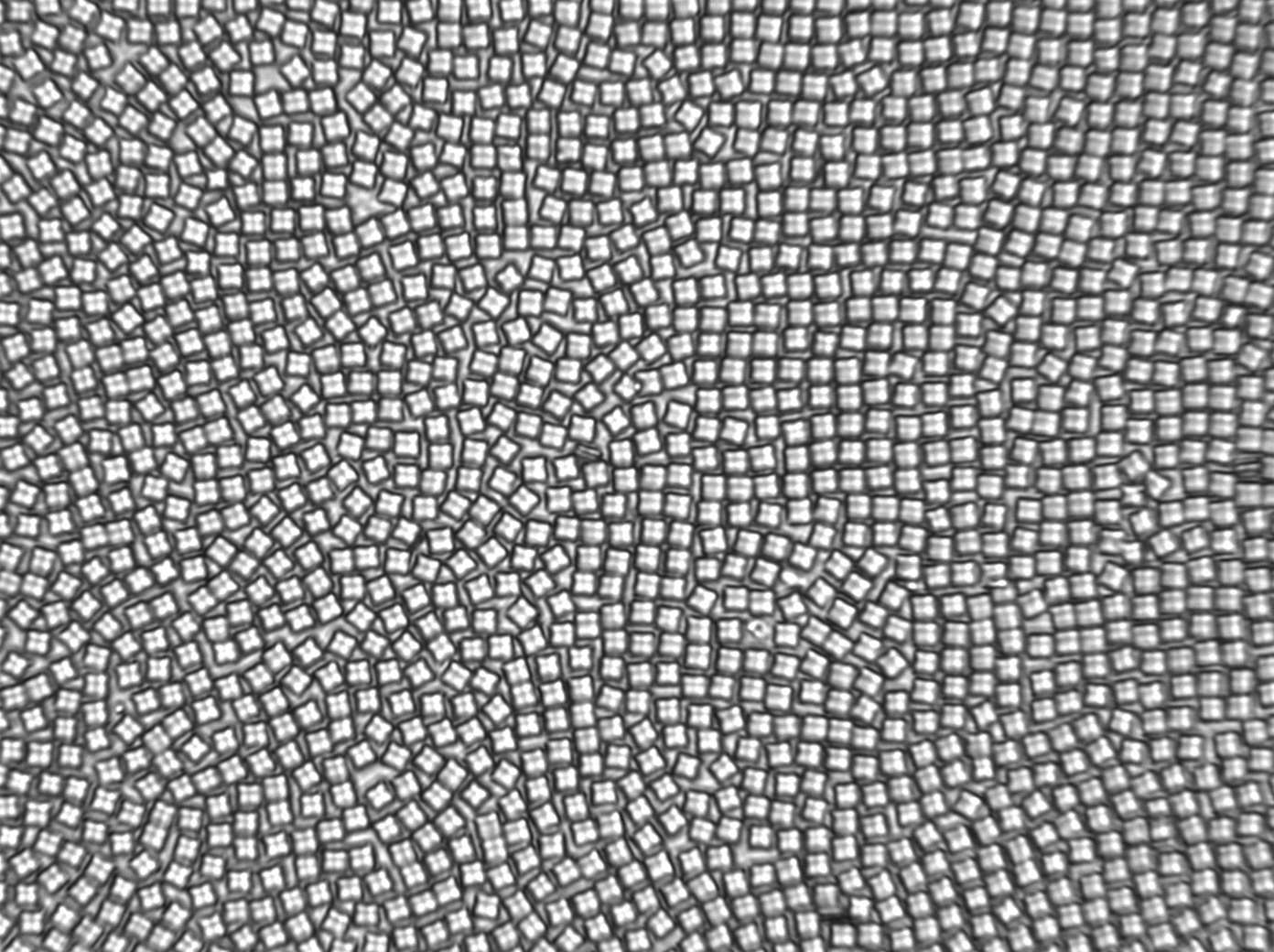}
  \caption{Micrograph of the 2D ensemble of cubes on a solid substrate where the curvature of the substrate can be controlled. The field of view typically contains between 900 to 1500 particles in the field of view of $210 \times 157~\mu\mathrm{m}^2$ imaged on $1392 \times 1040~\mathrm{pixel}^2$. 
  Here, the number of particles is $1350$ and the density. The ensemble of this example is in the tetratic phase (see below).}
  \label{fig:cubes}
\end{figure}

\begin{figure*}[t]
\centering
   \begin{minipage}[b]{.3\linewidth} 
      \includegraphics[width=\linewidth]{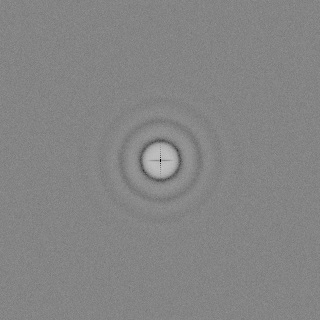}
   \end{minipage}
   \begin{minipage}[b]{.3\linewidth} 
      \includegraphics[width=\linewidth]{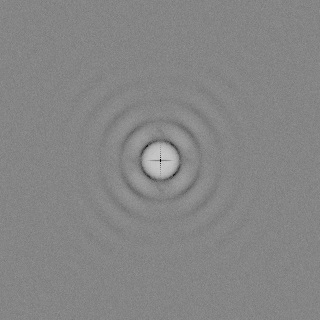}
   \end{minipage}
   \begin{minipage}[b]{.3\linewidth} 
     \includegraphics[width=\linewidth]{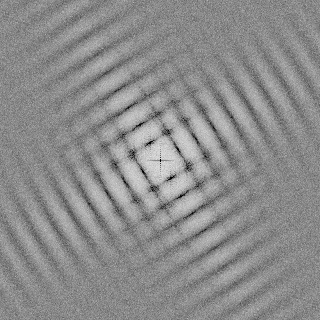}
   \end{minipage}
   \caption{Structure factor of the isotropic fluid (left plot), the hexatic phase (middle) and the crystalline phase (right).}
     \label{fig:structurefactor}
\end{figure*}

The colloidal solution is mounted in a sandwich geometry with a $80~\mu\textrm{m}$ thin cover slide in $2~\mathrm{mm}$ aqueous solution, formed using a teflon spacer ring, and sealed with a glass plate of $2~\mathrm{mm}$ thickness at the top. The thin, lower cover slide is glued with UV-glue (Norland Optical Adhesive 81) for sealing, while the upper glass plate is sealed with silicon rubber ring. The sandwich geometry is mounted in a copper block for thermal conductivity. The cell has a diameter of $20~\mathrm{mm}$ and is illuminated from below with an LED illumination (HLV2-22SW-1220-3W from CSS Inc.) and monitored with an CCD-camera from above (Marlin F-145B from Allied Vision) using an $20 \times$ objective and optical tube (Opto Sonderbedarf GmbH). The setup is similar to \cite{Ebert2009a}, with the exception being the lower interface water/air interface in hanging droplet geometry is a glass/air interface in the present work. The volume of the on average $2~\mathrm{mm}$ thick water layer can be adjusted with a syringe (Hamilton SYR 1mL 1001 TLL), driven by a micro-position system (Physical Instruments M-230.25). This way the thin lower glass plate can be curved from convex to concave with an amplitude up to $100~\mu\mathrm{m}$, where less than $10~\mu\mathrm{m}$ suffice for the present application. The aqueous solution contains $1~\%_{wt}$ Pluronics-F127 for sterical stabilization of the colloidal squares. Since the overlap volume of a flat surfaces is rather large, the steric stabilization works well and only a very small fraction of particles will be pinned on the solid glass surface. After mounting and sedimentation of the colloids, the 2D monolayer within the $20~\mathrm{mm}$ diameter cell is very dilute to avoid particles to form a multilayer after sedimentation. During a period of several weeks, the curvature of the confining interface is increased to become convex in steps of about $0.5~\mu\mathrm{m}$ and several days of waiting time. During this period, the particles sediment into the middle of the cell. Depending on the loss rate during writing, developing and washing the squares, a high concentrated colloidal monolayer in an area of about $5~\mathrm{mm}$ to $6~\mathrm{mm}$ in diameter can be achieved, with an area fraction between $0.5$ to $0.733$. Figure \ref{fig:setup} shows a sketch of the setup with the monolayer not to scale. The whole setup is mounted on a tripod with automated adjustable feet to keep the setup horizontal and the monolayer perpendicular to gravity with an accuracy of $\pm 10^{-6}~\mathrm{Rad}$ as in \cite{Ebert2009a}.\\

Figure \ref{fig:cubes} shows a micrograph of the colloidal ensemble. The field of view is $210 \times 157~\mathrm{\mu m}^2$ and contains $1350$ particles. Determined from Voronoi tesselation and neglecting particles less than $10~\mathrm{\mu m}$ to the border, the packing fraction is $0.679$. As will be shown later, the ensemble is in a tetratic phase at this density: particles are allowed to explore the plane by dislocation diffusion but the orientational degrees of freedom are restricted thus that the bond order correlation is quasi-long ranged.

\section{Image analysis}

\begin{figure*}[t]
\centering
   \begin{minipage}[b]{.3\linewidth} 
        \includegraphics[width=1.\textwidth]{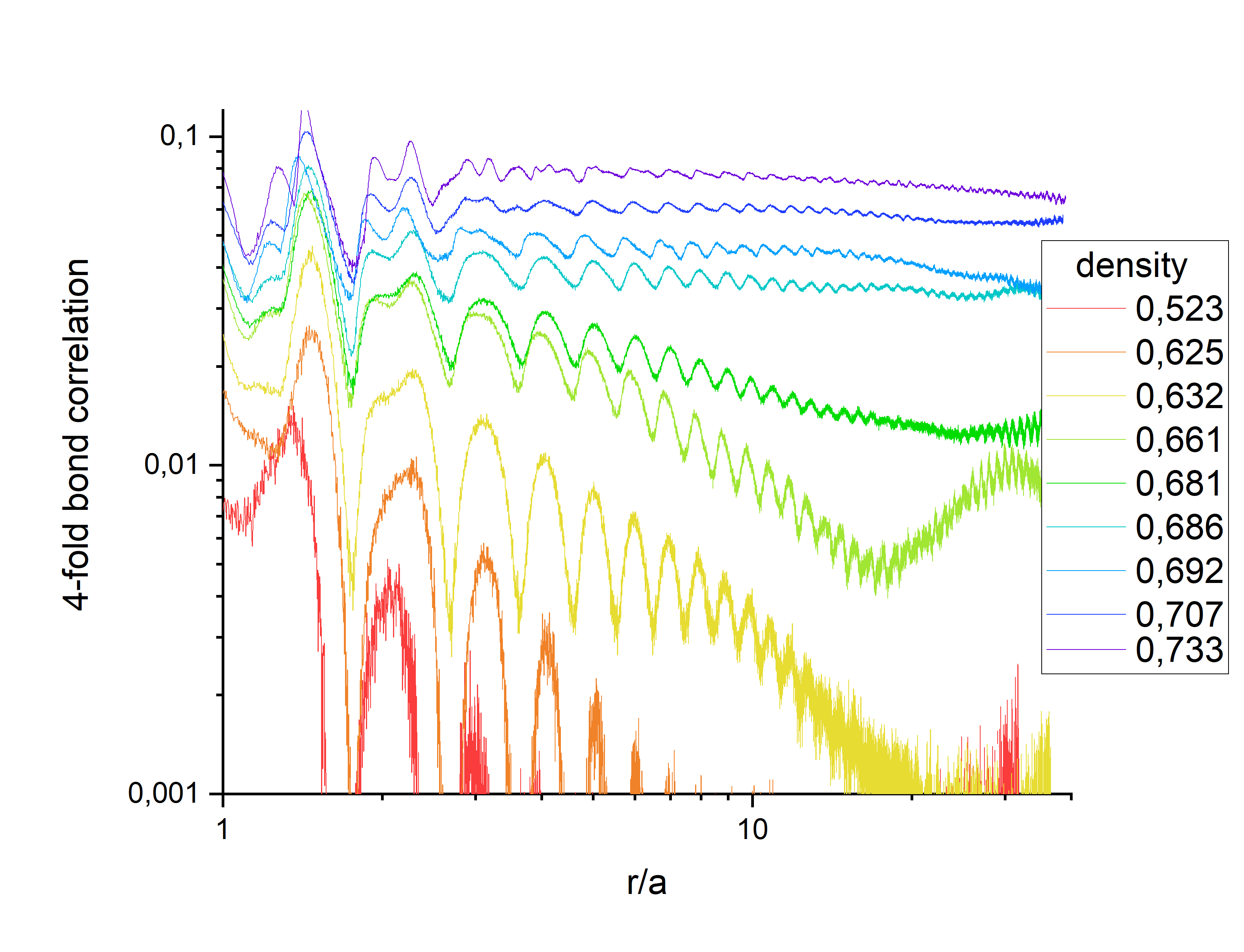}
   \end{minipage}
   \begin{minipage}[b]{.3\linewidth} 
        \includegraphics[width=1.\textwidth]{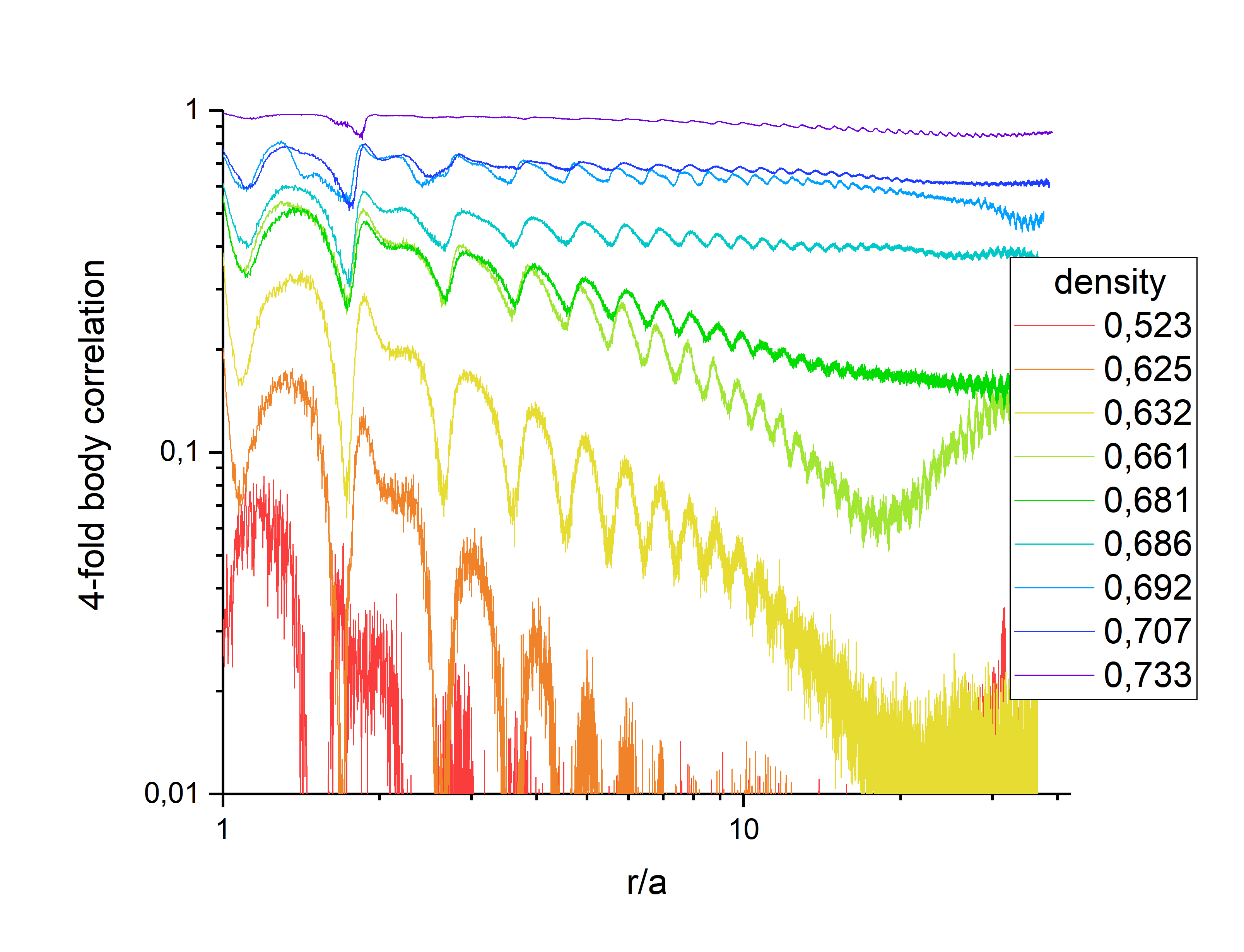}
   \end{minipage}
   \begin{minipage}[b]{.3\linewidth} 
        \includegraphics[width=1.\textwidth]{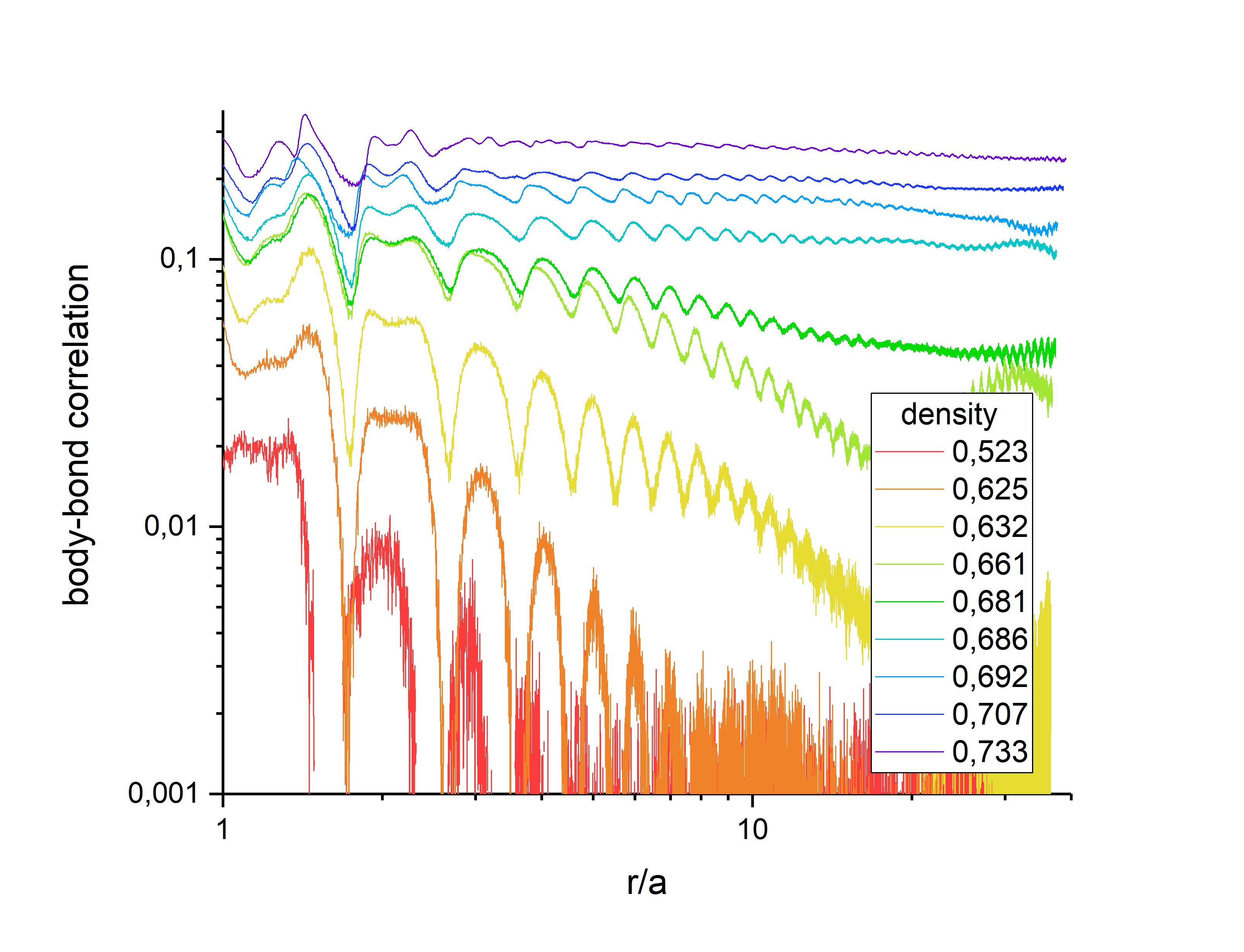}
   \end{minipage}
  \caption{Bond order correlation function $G_4(r)$ (left), body orientation correlation function $B_4(r)$ (middle), and bond-body cross correlation function $G_4B_4(r)$ (right) in log-log plots.}
    \label{fig:correlations}
\end{figure*}

The detection of particles is a variant of \cite{Ebert2009a} using the tracking software of John Crocker and David Grier for the detection of particle positions \cite{Crocker1996b}, written in IDL (Interactive Data Language)\footnote{https://www.nv5geospatialsoftware.com/Products/IDL}. Summarized, the image is binarized in black and white binary large objects (BLOBS). BLOBS are detected via connected-component analysis to identify individual colloidal particles. The orientation of the cubes is detected based on the method for detecting the local bond order field \cite{Keim2007} as follows \cite{Loeffler2018}: defining BLOB$_i$ to be the set of pixel that belong to the $i$th blob, we average over the pixel positions $\vec{r}_p$ with the 2D coordinates taken as complex number $\vec{r} = x + iy$ to calculate the center of mass
\begin{equation}
    \vec{r}_i = \langle \vec{r}_p \rangle_{p~\in~\mathrm{BLOB_i}} \quad .
\end{equation}
The center of mass of the BLOB serves as the particle position with sub-pixel resolution. The orientation $\xi_{4,i}$ of the $i$th squares is determined by looking at the pixel positions relative to the center of mass and analysing their distribution in four-fold space. The normalized orientation vector is
\begin{equation}
\label{eqn:xi}
    \xi_{4,i} = \tilde{\xi}_{4,i} / |\tilde{\xi}_{4,i}|
\end{equation}
with
\begin{equation}
    \tilde{\xi}_{4,i} = - \langle (\vec{r}_p - \vec{r}_i)^4 \rangle_{p~\in~\mathrm{BLOB_i}}
\end{equation}
and measures the body-orientation of the square. Note that the average of the four-fold pixel coordinates will point towards the corners of the cube, thus the negative value is taken so that $\xi_4$ shows the direction of the flat side of the square.\\

To save hard disc space, we do not record individual pictures but only size of the BLOBS as well as positional and body-orientational data of all $N$ particles in the field of view with a frame rate of $0.5~\mathrm{Hz}$ for a duration of up to $36$ hours at a given 2D density.

\section{Results}

The time-dependent particle positions are further processed to analyse the structure and phase of the colloidal ensemble. Here, we extract the structure factor and the bond- as well as body orientational correlation function to discriminate between isotropic fluid, tetratic and crystalline phase.

\subsection{Structure factor}

In close analogy to the search for the hexatic phase which shows up for isotropic particles \cite{Keim2007} and to discriminate the three phases and their symmetries, we are plotting the structure factor as calculated from the positional data for three different densities:
\begin{equation}
\label{sq}
 S(\vec{q}) = \frac{1}{N}\langle\sum_{\alpha,\alpha'}
e^{-i\vec{q}(\vec{r}_\alpha-\vec{r}_{\alpha'})}\rangle\quad.
\end{equation}
$\alpha, \alpha'$, run over all $N$ particles in the field of view and a time average is taken over $74$ configurations as shown in figure \ref{fig:structurefactor}. In the liquid phase, concentric rings appear having radii that can be connected to typical inter-particle distances. The tetratic phase is characterized by four segments of a ring which arise due to the quasi-long-range orientational order of the fourfold director. In the crystalline phase the Bragg peaks of finite width show up, reflecting the quasi-long-range character of the translational order in two dimensions due to Mermin-Wagner-Hohenberg fluctuations \cite{Mermin1966,Hohenberg1967,Mermin1968}. In principle one could check the azimuthal shape of the peaks for the tetratic versus crystalline phase: in the crystalline phase, the peak should have a Lorenzian form while in the tetratic phase, it should have the square root of a Lorenzian. Previous work has shown however that this criterion is numerically not very expressive, at least for the hexatic versus crystalline phase \cite{Dillmann2012a}.\\

\subsection{Bond- and body order correlations}

A more robust criterion is the decay of the bond-order correlation function $G_4(\vec{r})$, based on the four-fold local director field given by the nearest neighbours. Here the sum runs over the $N_j$ next neighbors of the particle $k$ at position $\vec{r}$ and $\theta_{jk}$ is the angle between a fixed reference axis and the bond of the particle $k$ and its neighbor $j$. $G_4(\vec{r})$ reads
\begin{equation}
\label{eqn:g4} G_4(r) = \langle \psi(\vec{r})\psi^\ast(\vec{0})
\rangle\quad ,
\end{equation}
with
\begin{equation}
\label{eqn:psi}
\psi(\vec{r}) = \psi_k = \frac{1}{N_j}\sum_{j}{e^{i4\theta_{jk}}}
\end{equation}
being the local four-fold director field. The average in \ref{eqn:3g4} is not only the ensemble average which is taken over all $N(N-1)/2$ particle-pair distances for each configuration but also averaged over $720$ statistically independent configurations. Using the bond-order correlation function, all three thermodynamic phases can be identified. For square crystals it follows that
\begin{eqnarray}
\label{eqn:3g4} \lim_{r\to\infty} &G_4(r)& \neq 0 \quad\quad\quad\
\mbox{\small{crystal: long range order}}\nonumber\\
&G_4(r)& \sim r^{-\eta_4} ~~\:\quad\mbox{\small{tetratic: quasi long range order}} \nonumber\\
&G_4(r)& \sim e^{-r/\xi_4} \:\quad\mbox{\small{isotropic: short
range order}}\nonumber\quad .
\end{eqnarray}
Figure \ref{fig:correlations} (left) shows $G_4(\vec{r})$ in a log-log plot, to highlight the algebraic decay. Qualitatively it behaves similar to colloidal ensembles with isotropic interaction, where the low symmetry phase is hexagonal \cite{Keim2007}. While a high particle density (purple curve) does not show a decay and represents the square crystal, a medium density (green curve) shows an algebraic decay for the tetratic phase, and low densities (orange/red curves) show exponential decay. $G_4(\vec{r})$ of an medium particle density (bright green curve) increases again with larger distances, which is explained by critical fluctuations: In the vicinity of the transition density, patches of symmetry broken domains become the size of the field of view. Only for infinitely large systems, the ensemble will be undoubtedly defined by its symmetry.\\

Quantitative results differ however from hexagonal crystals, where the amplitude of $G_6(\vec{r}=0) \approx 0.9$ deep in the crystal phase, while $G_4(\vec{r}=0) < 0.1$. This is due to determining the nearest neighbours via Voronoi tessellation: it is a trivalent network while at zero temperature, a tetravalent network fits best to a perfect square crystal. For any nonzero temperature, a tetravalent vertex will degenerate into two trivalent vertices due to thermal fluctuation. On average, the Voronoi tessellation will serve six nearest neighbours for any square crystal at finite temperatures, explaining why the normalization factor $N_j$ in Equation \ref{eqn:psi} is on average six instead of four. Furthermore, while four neighbours appear under $90^\circ$ and add up constructively in the four-fold director, two appear under multiples of $45^\circ$ and will be subtracted. Both effects effectively reduce the amplitude of $G_4(\vec{r})$ when nearest neighbours are determined by Voronoi tessellation. For hexagonal networks Voronoi tessellation works best and determines neighbours without doubt. Since the distance of neighbours under $45^\circ$ is about $\sqrt{2}$ longer, one can introduce a cut-off value for determining the four nearest neighbours, but this might interfere with correctly determining three- and five-fold dislocations. Another possibility is to weight the contribution of neighbours due to their part how they contribute to the circumference of the Voronoi-cell since the particles under multiples of $45^\circ$ contribute much less. Furthermore, a 2D variant of determining neighbours in a parameter free way as introduced in \cite{Meel2012} can be used. Determining which variant works best and fastest is a topic of ongoing work.\\

To check whether a rotator crystal shows up, where particles sit on lattice sites but are allowed to rotate relative to lattice directions, we have additionally investigated the body-body orientational correlation function. Based on Equation \ref{eqn:xi} which measures the body orientation along the edge of the square in vector of length unity, one can define the body correlation $B_4(\vec{r})$ in close analogy to Equation \ref{eqn:g4} as
\begin{equation}
\label{eqn:b4} B_4(r) = \langle \xi(\vec{r})\xi^\ast(\vec{0})
\rangle\quad .
\end{equation}
Figure \ref{fig:correlations} (middle) shows $B_4(\vec{r})$ in a log-log plot. Beside the amplitude which is $\approx 1$ for a crystal, the qualitative behavior is the same as for $G_4(\vec{r})$. This includes the density $\sigma = 0.661$ where critical behaviour is visible. The picture does not change taking the cross correlation of bond and body orientation as shown in Figure \ref{fig:correlations} (right) with
\begin{equation}
\label{eqn:g4b4}
    G_4B_4(\vec{r}) =  \langle \frac{1}{2} [\xi(\vec{r})\psi^\ast(\vec{0}) + \psi(\vec{r})\xi^\ast(\vec{0})] \rangle \quad .
\end{equation}
Bond and body orientation are highly correlated for all densities under investigation. A rotator crystal can be ruled out in our ensemble with particles written by 3D photo lithography. With this technology edges seem to be sharp enough also to avoid rhombic crystals.

\section{Conclusion}

Based on a 2D crystal of mesoscopic particles with square shape, the tetratic phase is validated experimentally in a thermal ensemble by analysing the structure factor and the four-fold bond-order correlation function: isotropic rings in $S(q)$ and an exponential decay of $G_4(r)$ mark the isotropic phase, a squared pattern in $S(q)$ and a not decaying $G_4(r)$ represents the 2D square crystal, while four segments of rings in $S(q)$ and an algebraic decay of $G_4(r)$ measures the tetratic phase. As seen in the structure factor, no rhombic phase is observed for all densities under investigation. Checking the body-orientational correlations and the bond-body cross correlation we did furthermore not observe a rotator crystal.

\section*{Author contributions}

R.L.: data curation, formal analysis, software, investigation. L.S.: data curation, formal analysis, software, investigation. P.K. formal analysis, software, project administration, supervision, writing.\\

\section*{Conflicts of interest}

There are no conflicts to declare.\\

\section*{Data availability}

The data-sets generated and analysed during the current study are available upon reasonable request by contacting the corresponding author.\\

\begin{acknowledgments}
L.S. acknowledges financial support from the Department of Physics, University of Konstanz. P.K. acknowledges support from the German Research Foundation (DFG) project number 453041792 (Heisenberg funding) and from SFB-1214 project B2. All authors acknowledge fruitful discussions with Georg Maret.
\end{acknowledgments}

\bibliography{literatur.bib}

\begin{thebibliography}{42}%
\makeatletter
\providecommand \@ifxundefined [1]{%
 \@ifx{#1\undefined}
}%
\providecommand \@ifnum [1]{%
 \ifnum #1\expandafter \@firstoftwo
 \else \expandafter \@secondoftwo
 \fi
}%
\providecommand \@ifx [1]{%
 \ifx #1\expandafter \@firstoftwo
 \else \expandafter \@secondoftwo
 \fi
}%
\providecommand \natexlab [1]{#1}%
\providecommand \enquote  [1]{``#1''}%
\providecommand \bibnamefont  [1]{#1}%
\providecommand \bibfnamefont [1]{#1}%
\providecommand \citenamefont [1]{#1}%
\providecommand \href@noop [0]{\@secondoftwo}%
\providecommand \href [0]{\begingroup \@sanitize@url \@href}%
\providecommand \@href[1]{\@@startlink{#1}\@@href}%
\providecommand \@@href[1]{\endgroup#1\@@endlink}%
\providecommand \@sanitize@url [0]{\catcode `\\12\catcode `\$12\catcode
  `\&12\catcode `\#12\catcode `\^12\catcode `\_12\catcode `\%12\relax}%
\providecommand \@@startlink[1]{}%
\providecommand \@@endlink[0]{}%
\providecommand \url  [0]{\begingroup\@sanitize@url \@url }%
\providecommand \@url [1]{\endgroup\@href {#1}{\urlprefix }}%
\providecommand \urlprefix  [0]{URL }%
\providecommand \Eprint [0]{\href }%
\providecommand \doibase [0]{https://doi.org/}%
\providecommand \selectlanguage [0]{\@gobble}%
\providecommand \bibinfo  [0]{\@secondoftwo}%
\providecommand \bibfield  [0]{\@secondoftwo}%
\providecommand \translation [1]{[#1]}%
\providecommand \BibitemOpen [0]{}%
\providecommand \bibitemStop [0]{}%
\providecommand \bibitemNoStop [0]{.\EOS\space}%
\providecommand \EOS [0]{\spacefactor3000\relax}%
\providecommand \BibitemShut  [1]{\csname bibitem#1\endcsname}%
\let\auto@bib@innerbib\@empty
\bibitem [{\citenamefont {Kosterlitz}\ and\ \citenamefont
  {Thouless}(1972)}]{Kosterlitz1972}%
  \BibitemOpen
  \bibfield  {author} {\bibinfo {author} {\bibfnamefont {J.~M.}\ \bibnamefont
  {Kosterlitz}}\ and\ \bibinfo {author} {\bibfnamefont {D.~J.}\ \bibnamefont
  {Thouless}},\ }\bibfield  {title} {\bibinfo {title} {Long range order and
  metastability in two dimensional solids and superfluids. (application of
  dislocation theory)},\ }\href {https://doi.org/10.1088/0022-3719/5/11/002}
  {\bibfield  {journal} {\bibinfo  {journal} {Journal of Physics C: Solid State
  Physics}\ }\textbf {\bibinfo {volume} {5}},\ \bibinfo {pages} {L124}
  (\bibinfo {year} {1972})}\BibitemShut {NoStop}%
\bibitem [{\citenamefont {Kosterlitz}\ and\ \citenamefont
  {Thouless}(1973)}]{Kosterlitz1973}%
  \BibitemOpen
  \bibfield  {author} {\bibinfo {author} {\bibfnamefont {J.~M.}\ \bibnamefont
  {Kosterlitz}}\ and\ \bibinfo {author} {\bibfnamefont {D.~J.}\ \bibnamefont
  {Thouless}},\ }\bibfield  {title} {\bibinfo {title} {Ordering, metastability
  and phase transitions in two-dimensional systems},\ }\href
  {https://doi.org/10.1088/0022-3719/6/7/010} {\bibfield  {journal} {\bibinfo
  {journal} {Journal of Physics C: Solid State Physics}\ }\textbf {\bibinfo
  {volume} {6}},\ \bibinfo {pages} {1181} (\bibinfo {year} {1973})}\BibitemShut
  {NoStop}%
\bibitem [{\citenamefont {Nelson}\ and\ \citenamefont
  {Kosterlitz}(1977)}]{Nelson1977}%
  \BibitemOpen
  \bibfield  {author} {\bibinfo {author} {\bibfnamefont {D.~R.}\ \bibnamefont
  {Nelson}}\ and\ \bibinfo {author} {\bibfnamefont {J.~M.}\ \bibnamefont
  {Kosterlitz}},\ }\bibfield  {title} {\bibinfo {title} {Universal jump in
  superfluid density of 2-dimensional superfluids},\ }\href
  {https://doi.org/DOI 10.1103/PhysRevLett.39.1201} {\bibfield  {journal}
  {\bibinfo  {journal} {Phys. Rev. Lett.}\ }\textbf {\bibinfo {volume} {39}},\
  \bibinfo {pages} {1201} (\bibinfo {year} {1977})}\BibitemShut {NoStop}%
\bibitem [{\citenamefont {Halperin}\ and\ \citenamefont
  {Nelson}(1978)}]{Halperin1978}%
  \BibitemOpen
  \bibfield  {author} {\bibinfo {author} {\bibfnamefont {B.~I.}\ \bibnamefont
  {Halperin}}\ and\ \bibinfo {author} {\bibfnamefont {D.~R.}\ \bibnamefont
  {Nelson}},\ }\bibfield  {title} {\bibinfo {title} {Theory of two-dimensional
  melting},\ }\href {https://doi.org/10.1103/PhysRevLett.41.121} {\bibfield
  {journal} {\bibinfo  {journal} {Phys. Rev. Lett.}\ }\textbf {\bibinfo
  {volume} {41}},\ \bibinfo {pages} {121} (\bibinfo {year} {1978})}\BibitemShut
  {NoStop}%
\bibitem [{\citenamefont {Nelson}\ and\ \citenamefont
  {Halperin}(1979)}]{Nelson1979}%
  \BibitemOpen
  \bibfield  {author} {\bibinfo {author} {\bibfnamefont {D.~R.}\ \bibnamefont
  {Nelson}}\ and\ \bibinfo {author} {\bibfnamefont {B.~I.}\ \bibnamefont
  {Halperin}},\ }\bibfield  {title} {\bibinfo {title} {Dislocation-mediated
  melting in two dimensions},\ }\href {https://doi.org/DOI
  10.1103/PhysRevB.19.2457} {\bibfield  {journal} {\bibinfo  {journal} {Phys.
  Rev. B}\ }\textbf {\bibinfo {volume} {19}},\ \bibinfo {pages} {2457}
  (\bibinfo {year} {1979})}\BibitemShut {NoStop}%
\bibitem [{\citenamefont {Young}(1979)}]{Young1979}%
  \BibitemOpen
  \bibfield  {author} {\bibinfo {author} {\bibfnamefont {A.~P.}\ \bibnamefont
  {Young}},\ }\bibfield  {title} {\bibinfo {title} {Melting and the vector
  coulomb gas in two dimensions},\ }\href {https://doi.org/DOI
  10.1103/PhysRevB.19.1855} {\bibfield  {journal} {\bibinfo  {journal} {Phys.
  Rev. B}\ }\textbf {\bibinfo {volume} {19}},\ \bibinfo {pages} {1855}
  (\bibinfo {year} {1979})}\BibitemShut {NoStop}%
\bibitem [{\citenamefont {Jaster}(1998)}]{Jaster1998}%
  \BibitemOpen
  \bibfield  {author} {\bibinfo {author} {\bibfnamefont {A.}~\bibnamefont
  {Jaster}},\ }\bibfield  {title} {\bibinfo {title} {Orientational order of the
  two-dimensional hard-disk system},\ }\href {https://doi.org/DOI
  10.1209/epl/i1998-00242-8} {\bibfield  {journal} {\bibinfo  {journal}
  {Europhy. Lett.}\ }\textbf {\bibinfo {volume} {42}},\ \bibinfo {pages} {277}
  (\bibinfo {year} {1998})}\BibitemShut {NoStop}%
\bibitem [{\citenamefont {Mak}(2006)}]{Mak2006}%
  \BibitemOpen
  \bibfield  {author} {\bibinfo {author} {\bibfnamefont {C.~H.}\ \bibnamefont
  {Mak}},\ }\bibfield  {title} {\bibinfo {title} {Large-scale simulations of
  the two-dimensional melting of hard disks},\ }\href {https://doi.org/Doi
  10.1103/Physreve.73.065104} {\bibfield  {journal} {\bibinfo  {journal} {Phys.
  Rev. E}\ }\textbf {\bibinfo {volume} {73}},\ \bibinfo {pages} {065104}
  (\bibinfo {year} {2006})}\BibitemShut {NoStop}%
\bibitem [{\citenamefont {Lin}\ \emph {et~al.}(2006)\citenamefont {Lin},
  \citenamefont {Zheng},\ and\ \citenamefont {Trimper}}]{Lin2006}%
  \BibitemOpen
  \bibfield  {author} {\bibinfo {author} {\bibfnamefont {S.~Z.}\ \bibnamefont
  {Lin}}, \bibinfo {author} {\bibfnamefont {B.}~\bibnamefont {Zheng}},\ and\
  \bibinfo {author} {\bibfnamefont {S.}~\bibnamefont {Trimper}},\ }\bibfield
  {title} {\bibinfo {title} {Computer simulations of two-dimensional melting
  with dipole-dipole interactions},\ }\href {https://doi.org/Doi
  10.1103/Physreve.73.066106} {\bibfield  {journal} {\bibinfo  {journal} {Phys.
  Rev. E}\ }\textbf {\bibinfo {volume} {73}},\ \bibinfo {pages} {066106}
  (\bibinfo {year} {2006})}\BibitemShut {NoStop}%
\bibitem [{\citenamefont {Murray}\ and\ \citenamefont
  {Van~Winkle}(1987)}]{Murray1987}%
  \BibitemOpen
  \bibfield  {author} {\bibinfo {author} {\bibfnamefont {C.~A.}\ \bibnamefont
  {Murray}}\ and\ \bibinfo {author} {\bibfnamefont {D.~H.}\ \bibnamefont
  {Van~Winkle}},\ }\bibfield  {title} {\bibinfo {title} {Experimental
  observation of two-stage melting in a classical two-dimensional screened
  coulomb system},\ }\href
  {https://journals.aps.org/prl/abstract/10.1103/PhysRevLett.58.1200}
  {\bibfield  {journal} {\bibinfo  {journal} {Phys. Rev. Lett.}\ }\textbf
  {\bibinfo {volume} {58}},\ \bibinfo {pages} {1200} (\bibinfo {year}
  {1987})}\BibitemShut {NoStop}%
\bibitem [{\citenamefont {Tang}\ \emph {et~al.}(1989)\citenamefont {Tang},
  \citenamefont {Armstrong}, \citenamefont {Mockler},\ and\ \citenamefont
  {Osullivan}}]{Tang1989}%
  \BibitemOpen
  \bibfield  {author} {\bibinfo {author} {\bibfnamefont {Y.}~\bibnamefont
  {Tang}}, \bibinfo {author} {\bibfnamefont {A.~J.}\ \bibnamefont {Armstrong}},
  \bibinfo {author} {\bibfnamefont {R.~C.}\ \bibnamefont {Mockler}},\ and\
  \bibinfo {author} {\bibfnamefont {W.~J.}\ \bibnamefont {Osullivan}},\
  }\bibfield  {title} {\bibinfo {title} {Free-expansion melting of a colloidal
  monolayer},\ }\href {https://doi.org/DOI 10.1103/PhysRevLett.62.2401}
  {\bibfield  {journal} {\bibinfo  {journal} {Phys. Rev. Lett.}\ }\textbf
  {\bibinfo {volume} {62}},\ \bibinfo {pages} {2401} (\bibinfo {year}
  {1989})}\BibitemShut {NoStop}%
\bibitem [{\citenamefont {Kusner}\ \emph {et~al.}(1994)\citenamefont {Kusner},
  \citenamefont {Mann}, \citenamefont {Kerins},\ and\ \citenamefont
  {Dahm}}]{Kusner1994}%
  \BibitemOpen
  \bibfield  {author} {\bibinfo {author} {\bibfnamefont {R.~E.}\ \bibnamefont
  {Kusner}}, \bibinfo {author} {\bibfnamefont {J.~A.}\ \bibnamefont {Mann}},
  \bibinfo {author} {\bibfnamefont {J.}~\bibnamefont {Kerins}},\ and\ \bibinfo
  {author} {\bibfnamefont {A.~J.}\ \bibnamefont {Dahm}},\ }\bibfield  {title}
  {\bibinfo {title} {Two-stage melting of a two-dimensional collodial lattice
  with dipole interactions},\ }\href
  {https://link.aps.org/doi/10.1103/PhysRevLett.73.3113} {\bibfield  {journal}
  {\bibinfo  {journal} {Phys. Rev. Lett.}\ }\textbf {\bibinfo {volume} {73}},\
  \bibinfo {pages} {3113} (\bibinfo {year} {1994})}\BibitemShut {NoStop}%
\bibitem [{\citenamefont {Kusner}\ \emph {et~al.}(1995)\citenamefont {Kusner},
  \citenamefont {Mann},\ and\ \citenamefont {Dahm}}]{Kusner1995}%
  \BibitemOpen
  \bibfield  {author} {\bibinfo {author} {\bibfnamefont {R.~E.}\ \bibnamefont
  {Kusner}}, \bibinfo {author} {\bibfnamefont {J.~A.}\ \bibnamefont {Mann}},\
  and\ \bibinfo {author} {\bibfnamefont {A.~J.}\ \bibnamefont {Dahm}},\
  }\bibfield  {title} {\bibinfo {title} {Two-stage melting in two dimensions in
  a system with dipole interactions},\ }\href
  {https://doi.org/10.1103/PhysRevB.51.5746} {\bibfield  {journal} {\bibinfo
  {journal} {Physical Review B}\ }\textbf {\bibinfo {volume} {51}},\ \bibinfo
  {pages} {5746} (\bibinfo {year} {1995})}\BibitemShut {NoStop}%
\bibitem [{\citenamefont {Marcus}\ and\ \citenamefont
  {Rice}(1996)}]{Marcus1996}%
  \BibitemOpen
  \bibfield  {author} {\bibinfo {author} {\bibfnamefont {A.~H.}\ \bibnamefont
  {Marcus}}\ and\ \bibinfo {author} {\bibfnamefont {S.~A.}\ \bibnamefont
  {Rice}},\ }\bibfield  {title} {\bibinfo {title} {Observations of first-order
  liquid-to-hexatic and hexatic-to-solid phase transitions in a confined
  colloid suspension},\ }\href
  {https://doi.org/https://doi.org/10.1103/PhysRevLett.77.2577} {\bibfield
  {journal} {\bibinfo  {journal} {Phys. Rev. Lett.}\ }\textbf {\bibinfo
  {volume} {77}},\ \bibinfo {pages} {2577} (\bibinfo {year}
  {1996})}\BibitemShut {NoStop}%
\bibitem [{\citenamefont {Zahn}\ \emph {et~al.}(1999)\citenamefont {Zahn},
  \citenamefont {Lenke},\ and\ \citenamefont {Maret}}]{Zahn1999}%
  \BibitemOpen
  \bibfield  {author} {\bibinfo {author} {\bibfnamefont {K.}~\bibnamefont
  {Zahn}}, \bibinfo {author} {\bibfnamefont {R.}~\bibnamefont {Lenke}},\ and\
  \bibinfo {author} {\bibfnamefont {G.}~\bibnamefont {Maret}},\ }\bibfield
  {title} {\bibinfo {title} {Two-stage melting of paramagnetic colloidal
  crystals in two dimensions},\ }\href {https://doi.org/DOI
  10.1103/PhysRevLett.82.2721} {\bibfield  {journal} {\bibinfo  {journal}
  {Phys. Rev. Lett.}\ }\textbf {\bibinfo {volume} {82}},\ \bibinfo {pages}
  {2721} (\bibinfo {year} {1999})}\BibitemShut {NoStop}%
\bibitem [{\citenamefont {Zahn}\ and\ \citenamefont {Maret}(2000)}]{Zahn2000}%
  \BibitemOpen
  \bibfield  {author} {\bibinfo {author} {\bibfnamefont {K.}~\bibnamefont
  {Zahn}}\ and\ \bibinfo {author} {\bibfnamefont {G.}~\bibnamefont {Maret}},\
  }\bibfield  {title} {\bibinfo {title} {Dynamic criteria for melting in two
  dimensions},\ }\href {https://doi.org/DOI 10.1103/PhysRevLett.85.3656}
  {\bibfield  {journal} {\bibinfo  {journal} {Phys. Rev. Lett.}\ }\textbf
  {\bibinfo {volume} {85}},\ \bibinfo {pages} {3656} (\bibinfo {year}
  {2000})}\BibitemShut {NoStop}%
\bibitem [{\citenamefont {Eisenmann}\ \emph {et~al.}(2005)\citenamefont
  {Eisenmann}, \citenamefont {Gasser}, \citenamefont {Keim}, \citenamefont
  {Maret},\ and\ \citenamefont {von Grünberg}}]{Eisenmann2005}%
  \BibitemOpen
  \bibfield  {author} {\bibinfo {author} {\bibfnamefont {C.}~\bibnamefont
  {Eisenmann}}, \bibinfo {author} {\bibfnamefont {U.}~\bibnamefont {Gasser}},
  \bibinfo {author} {\bibfnamefont {P.}~\bibnamefont {Keim}}, \bibinfo {author}
  {\bibfnamefont {G.}~\bibnamefont {Maret}},\ and\ \bibinfo {author}
  {\bibfnamefont {H.~H.}\ \bibnamefont {von Grünberg}},\ }\bibfield  {title}
  {\bibinfo {title} {Pair interaction of dislocations in two-dimensional
  crystals},\ }\href
  {https://doi.org/https://doi.org/10.1103/PhysRevLett.95.185502} {\bibfield
  {journal} {\bibinfo  {journal} {Phys. Rev. Lett.}\ }\textbf {\bibinfo
  {volume} {95}},\ \bibinfo {pages} {185502} (\bibinfo {year}
  {2005})}\BibitemShut {NoStop}%
\bibitem [{\citenamefont {Zanghellini}\ \emph {et~al.}(2005)\citenamefont
  {Zanghellini}, \citenamefont {Keim},\ and\ \citenamefont
  {Grünberg}}]{Zanghellini2005}%
  \BibitemOpen
  \bibfield  {author} {\bibinfo {author} {\bibfnamefont {J.}~\bibnamefont
  {Zanghellini}}, \bibinfo {author} {\bibfnamefont {P.}~\bibnamefont {Keim}},\
  and\ \bibinfo {author} {\bibfnamefont {H.~H.~v.}\ \bibnamefont {Grünberg}},\
  }\bibfield  {title} {\bibinfo {title} {The softening of two-dimensional
  colloidal crystals},\ }\href
  {https://doi.org/https://doi.org/10.1088/0953-8984/17/45/051} {\bibfield
  {journal} {\bibinfo  {journal} {J. Phys. Cond. Mat.}\ }\textbf {\bibinfo
  {volume} {17}},\ \bibinfo {pages} {S3579} (\bibinfo {year}
  {2005})}\BibitemShut {NoStop}%
\bibitem [{\citenamefont {Keim}\ \emph {et~al.}(2007)\citenamefont {Keim},
  \citenamefont {Maret},\ and\ \citenamefont {von Grünberg}}]{Keim2007}%
  \BibitemOpen
  \bibfield  {author} {\bibinfo {author} {\bibfnamefont {P.}~\bibnamefont
  {Keim}}, \bibinfo {author} {\bibfnamefont {G.}~\bibnamefont {Maret}},\ and\
  \bibinfo {author} {\bibfnamefont {H.~H.}\ \bibnamefont {von Grünberg}},\
  }\bibfield  {title} {\bibinfo {title} {Frank's constant in the hexatic
  phase},\ }\href {https://doi.org/https://doi.org/10.1103/PhysRevE.75.031402}
  {\bibfield  {journal} {\bibinfo  {journal} {Phys. Rev. E.}\ }\textbf
  {\bibinfo {volume} {75}},\ \bibinfo {pages} {031402} (\bibinfo {year}
  {2007})}\BibitemShut {NoStop}%
\bibitem [{\citenamefont {Gasser}\ \emph {et~al.}(2010)\citenamefont {Gasser},
  \citenamefont {Eisenmann}, \citenamefont {Maret},\ and\ \citenamefont
  {Keim}}]{Gasser2010}%
  \BibitemOpen
  \bibfield  {author} {\bibinfo {author} {\bibfnamefont {U.}~\bibnamefont
  {Gasser}}, \bibinfo {author} {\bibfnamefont {C.}~\bibnamefont {Eisenmann}},
  \bibinfo {author} {\bibfnamefont {G.}~\bibnamefont {Maret}},\ and\ \bibinfo
  {author} {\bibfnamefont {P.}~\bibnamefont {Keim}},\ }\bibfield  {title}
  {\bibinfo {title} {Melting of crystals in two dimensions},\ }\href
  {https://doi.org/https://doi.org/10.1002/cphc.200900755} {\bibfield
  {journal} {\bibinfo  {journal} {ChemPhysChem}\ }\textbf {\bibinfo {volume}
  {11}},\ \bibinfo {pages} {963} (\bibinfo {year} {2010})}\BibitemShut
  {NoStop}%
\bibitem [{\citenamefont {Thorneywork}\ \emph {et~al.}(2017)\citenamefont
  {Thorneywork}, \citenamefont {Abbott}, \citenamefont {Aarts},\ and\
  \citenamefont {Dullens}}]{Thorneywork2017}%
  \BibitemOpen
  \bibfield  {author} {\bibinfo {author} {\bibfnamefont {A.~L.}\ \bibnamefont
  {Thorneywork}}, \bibinfo {author} {\bibfnamefont {J.~L.}\ \bibnamefont
  {Abbott}}, \bibinfo {author} {\bibfnamefont {D.~G. A.~L.}\ \bibnamefont
  {Aarts}},\ and\ \bibinfo {author} {\bibfnamefont {R.~P.~A.}\ \bibnamefont
  {Dullens}},\ }\bibfield  {title} {\bibinfo {title} {Two-dimensional melting
  of colloidal hard spheres},\ }\href
  {https://link.aps.org/doi/10.1103/PhysRevLett.118.158001} {\bibfield
  {journal} {\bibinfo  {journal} {Physical Review Letters}\ }\textbf {\bibinfo
  {volume} {118}},\ \bibinfo {pages} {158001} (\bibinfo {year}
  {2017})}\BibitemShut {NoStop}%
\bibitem [{\citenamefont {Thorneywork}\ \emph {et~al.}(2018)\citenamefont
  {Thorneywork}, \citenamefont {Abbott}, \citenamefont {Aarts}, \citenamefont
  {Keim},\ and\ \citenamefont {Dullens}}]{Thorneywork2018}%
  \BibitemOpen
  \bibfield  {author} {\bibinfo {author} {\bibfnamefont {A.~L.}\ \bibnamefont
  {Thorneywork}}, \bibinfo {author} {\bibfnamefont {J.~L.}\ \bibnamefont
  {Abbott}}, \bibinfo {author} {\bibfnamefont {D.~G. A.~L.}\ \bibnamefont
  {Aarts}}, \bibinfo {author} {\bibfnamefont {P.}~\bibnamefont {Keim}},\ and\
  \bibinfo {author} {\bibfnamefont {R.~P.~A.}\ \bibnamefont {Dullens}},\
  }\bibfield  {title} {\bibinfo {title} {Bond-orientational order and frank’s
  constant in two-dimensional colloidal hard spheres},\ }\href
  {https://doi.org/https://doi.org/10.1088/1361-648X/aaab31} {\bibfield
  {journal} {\bibinfo  {journal} {Journal of Physics: Condensed Matter}\
  }\textbf {\bibinfo {volume} {30}},\ \bibinfo {pages} {104003} (\bibinfo
  {year} {2018})}\BibitemShut {NoStop}%
\bibitem [{\citenamefont {Bernard}\ and\ \citenamefont
  {Krauth}(2011)}]{Bernard2011}%
  \BibitemOpen
  \bibfield  {author} {\bibinfo {author} {\bibfnamefont {E.~P.}\ \bibnamefont
  {Bernard}}\ and\ \bibinfo {author} {\bibfnamefont {W.}~\bibnamefont
  {Krauth}},\ }\bibfield  {title} {\bibinfo {title} {Two-step melting in two
  dimensions: First-order liquid-hexatic transition},\ }\href
  {http://link.aps.org/doi/10.1103/PhysRevLett.107.155704} {\bibfield
  {journal} {\bibinfo  {journal} {Phys. Rev. Lett.}\ }\textbf {\bibinfo
  {volume} {107}},\ \bibinfo {pages} {155704} (\bibinfo {year}
  {2011})}\BibitemShut {NoStop}%
\bibitem [{\citenamefont {Engel}\ \emph {et~al.}(2013)\citenamefont {Engel},
  \citenamefont {Anderson}, \citenamefont {Glotzer}, \citenamefont {Isobe},
  \citenamefont {Bernard},\ and\ \citenamefont {Krauth}}]{Engel2013}%
  \BibitemOpen
  \bibfield  {author} {\bibinfo {author} {\bibfnamefont {M.}~\bibnamefont
  {Engel}}, \bibinfo {author} {\bibfnamefont {J.~A.}\ \bibnamefont {Anderson}},
  \bibinfo {author} {\bibfnamefont {S.~C.}\ \bibnamefont {Glotzer}}, \bibinfo
  {author} {\bibfnamefont {M.}~\bibnamefont {Isobe}}, \bibinfo {author}
  {\bibfnamefont {E.~P.}\ \bibnamefont {Bernard}},\ and\ \bibinfo {author}
  {\bibfnamefont {W.}~\bibnamefont {Krauth}},\ }\bibfield  {title} {\bibinfo
  {title} {Hard-disk equation of state: First-order liquid-hexatic transition
  in two dimensions with three simulation methods},\ }\href
  {https://doi.org/10.1103/Physreve.87.042134} {\bibfield  {journal} {\bibinfo
  {journal} {Phys. Rev. E}\ }\textbf {\bibinfo {volume} {87}},\ \bibinfo
  {pages} {042134} (\bibinfo {year} {2013})}\BibitemShut {NoStop}%
\bibitem [{\citenamefont {Kapfer}\ and\ \citenamefont
  {Krauth}(2015)}]{Kapfer2015}%
  \BibitemOpen
  \bibfield  {author} {\bibinfo {author} {\bibfnamefont {S.~C.}\ \bibnamefont
  {Kapfer}}\ and\ \bibinfo {author} {\bibfnamefont {W.}~\bibnamefont
  {Krauth}},\ }\bibfield  {title} {\bibinfo {title} {Two-dimensional melting:
  From liquid-hexatic coexistence to continuous transitions},\ }\href
  {http://link.aps.org/doi/10.1103/PhysRevLett.114.035702} {\bibfield
  {journal} {\bibinfo  {journal} {Phys. Rev. Lett.}\ }\textbf {\bibinfo
  {volume} {114}},\ \bibinfo {pages} {035702} (\bibinfo {year}
  {2015})}\BibitemShut {NoStop}%
\bibitem [{\citenamefont {Eisenmann}\ \emph
  {et~al.}(2004{\natexlab{a}})\citenamefont {Eisenmann}, \citenamefont
  {Gasser}, \citenamefont {Keim},\ and\ \citenamefont {Maret}}]{Eisenmann2004}%
  \BibitemOpen
  \bibfield  {author} {\bibinfo {author} {\bibfnamefont {C.}~\bibnamefont
  {Eisenmann}}, \bibinfo {author} {\bibfnamefont {U.}~\bibnamefont {Gasser}},
  \bibinfo {author} {\bibfnamefont {P.}~\bibnamefont {Keim}},\ and\ \bibinfo
  {author} {\bibfnamefont {G.}~\bibnamefont {Maret}},\ }\bibfield  {title}
  {\bibinfo {title} {Anisotropic defect-mediated melting of two-dimensional
  colloidal crystals},\ }\href
  {https://doi.org/https://doi.org/10.1103/PhysRevLett.93.105702} {\bibfield
  {journal} {\bibinfo  {journal} {Phys. Rev. Lett.}\ }\textbf {\bibinfo
  {volume} {93}},\ \bibinfo {pages} {105702} (\bibinfo {year}
  {2004}{\natexlab{a}})}\BibitemShut {NoStop}%
\bibitem [{\citenamefont {Eisenmann}\ \emph
  {et~al.}(2004{\natexlab{b}})\citenamefont {Eisenmann}, \citenamefont {Keim},
  \citenamefont {Gasser},\ and\ \citenamefont {Maret}}]{Eisenmann2004b}%
  \BibitemOpen
  \bibfield  {author} {\bibinfo {author} {\bibfnamefont {C.}~\bibnamefont
  {Eisenmann}}, \bibinfo {author} {\bibfnamefont {P.}~\bibnamefont {Keim}},
  \bibinfo {author} {\bibfnamefont {U.}~\bibnamefont {Gasser}},\ and\ \bibinfo
  {author} {\bibfnamefont {G.}~\bibnamefont {Maret}},\ }\bibfield  {title}
  {\bibinfo {title} {Melting of anisotropic colloidal crystals in two
  dimensions},\ }\href
  {https://doi.org/https://doi.org/10.1088/0953-8984/16/38/024} {\bibfield
  {journal} {\bibinfo  {journal} {Jour. Phys. Cond. Matt.}\ }\textbf {\bibinfo
  {volume} {16}},\ \bibinfo {pages} {4095} (\bibinfo {year}
  {2004}{\natexlab{b}})}\BibitemShut {NoStop}%
\bibitem [{\citenamefont {Froltsov}\ \emph {et~al.}(2005)\citenamefont
  {Froltsov}, \citenamefont {Likos}, \citenamefont {Löwen}, \citenamefont
  {Eisenmann}, \citenamefont {Gasser}, \citenamefont {Keim},\ and\
  \citenamefont {Maret}}]{Froltsov2005}%
  \BibitemOpen
  \bibfield  {author} {\bibinfo {author} {\bibfnamefont {V.~A.}\ \bibnamefont
  {Froltsov}}, \bibinfo {author} {\bibfnamefont {C.~N.}\ \bibnamefont {Likos}},
  \bibinfo {author} {\bibfnamefont {H.}~\bibnamefont {Löwen}}, \bibinfo
  {author} {\bibfnamefont {C.}~\bibnamefont {Eisenmann}}, \bibinfo {author}
  {\bibfnamefont {U.}~\bibnamefont {Gasser}}, \bibinfo {author} {\bibfnamefont
  {P.}~\bibnamefont {Keim}},\ and\ \bibinfo {author} {\bibfnamefont
  {G.}~\bibnamefont {Maret}},\ }\bibfield  {title} {\bibinfo {title}
  {Anisotropic mean-square displacements in two-dimensional colloidal crystals
  of tilted dipoles},\ }\href
  {https://doi.org/https://doi.org/10.1103/PhysRevE.71.031404} {\bibfield
  {journal} {\bibinfo  {journal} {Phys. Rev. E}\ }\textbf {\bibinfo {volume}
  {71}},\ \bibinfo {pages} {031404} (\bibinfo {year} {2005})}\BibitemShut
  {NoStop}%
\bibitem [{\citenamefont {Wojciechowski}\ and\ \citenamefont
  {Frenkel}(2004)}]{Wojciechowski2004}%
  \BibitemOpen
  \bibfield  {author} {\bibinfo {author} {\bibfnamefont {K.}~\bibnamefont
  {Wojciechowski}}\ and\ \bibinfo {author} {\bibfnamefont {D.}~\bibnamefont
  {Frenkel}},\ }\bibfield  {title} {\bibinfo {title} {Tetratic phase in the
  planar hard square system?},\ }\href
  {https://doi.org/10.12921/cmst.2004.10.02.235-255} {\bibfield  {journal}
  {\bibinfo  {journal} {CMST}\ }\textbf {\bibinfo {volume} {10}},\ \bibinfo
  {pages} {235} (\bibinfo {year} {2004})}\BibitemShut {NoStop}%
\bibitem [{\citenamefont {Smallenburg}\ \emph {et~al.}(2012)\citenamefont
  {Smallenburg}, \citenamefont {Filion}, \citenamefont {Marechal},\ and\
  \citenamefont {Dijkstra}}]{Smallenburg2012}%
  \BibitemOpen
  \bibfield  {author} {\bibinfo {author} {\bibfnamefont {F.}~\bibnamefont
  {Smallenburg}}, \bibinfo {author} {\bibfnamefont {L.}~\bibnamefont {Filion}},
  \bibinfo {author} {\bibfnamefont {M.}~\bibnamefont {Marechal}},\ and\
  \bibinfo {author} {\bibfnamefont {M.}~\bibnamefont {Dijkstra}},\ }\bibfield
  {title} {\bibinfo {title} {Vacancy-stabilized crystalline order in hard
  cubes},\ }\href {https://doi.org/DOI 10.1073/pnas.1211784109} {\bibfield
  {journal} {\bibinfo  {journal} {Proc. Natl. Acad. Sci.}\ }\textbf {\bibinfo
  {volume} {109}},\ \bibinfo {pages} {17886} (\bibinfo {year}
  {2012})}\BibitemShut {NoStop}%
\bibitem [{\citenamefont {Zhao}\ \emph {et~al.}(2011)\citenamefont {Zhao},
  \citenamefont {Bruinsma},\ and\ \citenamefont {Mason}}]{Zhao2011}%
  \BibitemOpen
  \bibfield  {author} {\bibinfo {author} {\bibfnamefont {K.}~\bibnamefont
  {Zhao}}, \bibinfo {author} {\bibfnamefont {R.}~\bibnamefont {Bruinsma}},\
  and\ \bibinfo {author} {\bibfnamefont {T.~G.}\ \bibnamefont {Mason}},\
  }\bibfield  {title} {\bibinfo {title} {Entropic crystal–crystal transitions
  of brownian squares},\ }\href {https://doi.org/10.1073/pnas.1014942108}
  {\bibfield  {journal} {\bibinfo  {journal} {Proceedings of the National
  Academy of Sciences}\ }\textbf {\bibinfo {volume} {108}},\ \bibinfo {pages}
  {2684} (\bibinfo {year} {2011})}\BibitemShut {NoStop}%
\bibitem [{\citenamefont {Avendano}\ and\ \citenamefont
  {Escobedo}(2012)}]{Avendano2012}%
  \BibitemOpen
  \bibfield  {author} {\bibinfo {author} {\bibfnamefont {C.}~\bibnamefont
  {Avendano}}\ and\ \bibinfo {author} {\bibfnamefont {F.~A.}\ \bibnamefont
  {Escobedo}},\ }\bibfield  {title} {\bibinfo {title} {Phase behavior of
  rounded hard-squares},\ }\href {https://doi.org/10.1039/C2SM07428A}
  {\bibfield  {journal} {\bibinfo  {journal} {Soft Matter}\ }\textbf {\bibinfo
  {volume} {8}},\ \bibinfo {pages} {4675} (\bibinfo {year} {2012})}\BibitemShut
  {NoStop}%
\bibitem [{\citenamefont {Donev}\ \emph {et~al.}(2006)\citenamefont {Donev},
  \citenamefont {Burton}, \citenamefont {Stillinger},\ and\ \citenamefont
  {Torquato}}]{Donev2006}%
  \BibitemOpen
  \bibfield  {author} {\bibinfo {author} {\bibfnamefont {A.}~\bibnamefont
  {Donev}}, \bibinfo {author} {\bibfnamefont {J.}~\bibnamefont {Burton}},
  \bibinfo {author} {\bibfnamefont {F.~H.}\ \bibnamefont {Stillinger}},\ and\
  \bibinfo {author} {\bibfnamefont {S.}~\bibnamefont {Torquato}},\ }\bibfield
  {title} {\bibinfo {title} {Tetratic order in the phase behavior of a
  hard-rectangle system},\ }\href {https://doi.org/10.1103/PhysRevB.73.054109}
  {\bibfield  {journal} {\bibinfo  {journal} {Physical Review B}\ }\textbf
  {\bibinfo {volume} {73}},\ \bibinfo {pages} {054109} (\bibinfo {year}
  {2006})},\ \bibinfo {note} {pRB}\BibitemShut {NoStop}%
\bibitem [{\citenamefont {Dertli}\ and\ \citenamefont
  {Speck}(2024)}]{Dertli2024}%
  \BibitemOpen
  \bibfield  {author} {\bibinfo {author} {\bibfnamefont {D.}~\bibnamefont
  {Dertli}}\ and\ \bibinfo {author} {\bibfnamefont {T.}~\bibnamefont {Speck}},\
  }\href {https://arxiv.org/abs/2408.06889} {\bibinfo {title} {In pursuit of
  the tetratic phase in hard rectangles}} (\bibinfo {year} {2024}),\ \Eprint
  {https://arxiv.org/abs/2408.06889} {arXiv:2408.06889 [cond-mat.soft]}
  \BibitemShut {NoStop}%
\bibitem [{\citenamefont {Ebert}\ \emph {et~al.}(2009)\citenamefont {Ebert},
  \citenamefont {Dillmann}, \citenamefont {Maret},\ and\ \citenamefont
  {Keim}}]{Ebert2009a}%
  \BibitemOpen
  \bibfield  {author} {\bibinfo {author} {\bibfnamefont {F.}~\bibnamefont
  {Ebert}}, \bibinfo {author} {\bibfnamefont {P.}~\bibnamefont {Dillmann}},
  \bibinfo {author} {\bibfnamefont {G.}~\bibnamefont {Maret}},\ and\ \bibinfo
  {author} {\bibfnamefont {P.}~\bibnamefont {Keim}},\ }\bibfield  {title}
  {\bibinfo {title} {The experimental realization of a two-dimensional
  colloidal model system},\ }\href
  {https://doi.org/https://doi.org/10.1063/1.3188948} {\bibfield  {journal}
  {\bibinfo  {journal} {Rev. Sci. Inst.}\ }\textbf {\bibinfo {volume} {80}},\
  \bibinfo {pages} {083902} (\bibinfo {year} {2009})}\BibitemShut {NoStop}%
\bibitem [{\citenamefont {Crocker}\ and\ \citenamefont
  {Grier}(1996)}]{Crocker1996b}%
  \BibitemOpen
  \bibfield  {author} {\bibinfo {author} {\bibfnamefont {J.~C.}\ \bibnamefont
  {Crocker}}\ and\ \bibinfo {author} {\bibfnamefont {D.~G.}\ \bibnamefont
  {Grier}},\ }\bibfield  {title} {\bibinfo {title} {Methods of digital video
  microscopy for colloidal studies},\ }\href {https://doi.org/DOI
  10.1006/jcis.1996.0217} {\bibfield  {journal} {\bibinfo  {journal} {J. Coll.
  Interf. Sci.}\ }\textbf {\bibinfo {volume} {179}},\ \bibinfo {pages} {298}
  (\bibinfo {year} {1996})}\BibitemShut {NoStop}%
\bibitem [{Note1()}]{Note1}%
  \BibitemOpen
  \bibinfo {note}
  {Https://www.nv5geospatialsoftware.com/Products/IDL}\BibitemShut {NoStop}%
\bibitem [{\citenamefont {Mermin}\ and\ \citenamefont
  {Wagner}(1966)}]{Mermin1966}%
  \BibitemOpen
  \bibfield  {author} {\bibinfo {author} {\bibfnamefont {N.~D.}\ \bibnamefont
  {Mermin}}\ and\ \bibinfo {author} {\bibfnamefont {H.}~\bibnamefont
  {Wagner}},\ }\bibfield  {title} {\bibinfo {title} {Absence of ferromagnetism
  or antiferromagnetism in one- or two-dimensional isotropic heisenberg
  models},\ }\href {https://doi.org/DOI 10.1103/PhysRevLett.17.1133} {\bibfield
   {journal} {\bibinfo  {journal} {Phys. Rev. Lett.}\ }\textbf {\bibinfo
  {volume} {17}},\ \bibinfo {pages} {1133} (\bibinfo {year}
  {1966})}\BibitemShut {NoStop}%
\bibitem [{\citenamefont {Hohenberg}(1967)}]{Hohenberg1967}%
  \BibitemOpen
  \bibfield  {author} {\bibinfo {author} {\bibfnamefont {P.}~\bibnamefont
  {Hohenberg}},\ }\bibfield  {title} {\bibinfo {title} {Existence of long-range
  order in 1 and 2 dimensions},\ }\href
  {https://doi.org/10.1103/PhysRev.158.383} {\bibfield  {journal} {\bibinfo
  {journal} {Phys. Rev.}\ }\textbf {\bibinfo {volume} {158}},\ \bibinfo {pages}
  {383} (\bibinfo {year} {1967})}\BibitemShut {NoStop}%
\bibitem [{\citenamefont {Mermin}(1968)}]{Mermin1968}%
  \BibitemOpen
  \bibfield  {author} {\bibinfo {author} {\bibfnamefont {N.~D.}\ \bibnamefont
  {Mermin}},\ }\bibfield  {title} {\bibinfo {title} {Crystalline order in two
  dimensions},\ }\href {https://doi.org/10.1103/PhysRev.176.250} {\bibfield
  {journal} {\bibinfo  {journal} {Phys. Rev.}\ }\textbf {\bibinfo {volume}
  {176}},\ \bibinfo {pages} {250} (\bibinfo {year} {1968})}\BibitemShut
  {NoStop}%
\bibitem [{\citenamefont {Dillmann}\ \emph {et~al.}(2012)\citenamefont
  {Dillmann}, \citenamefont {Maret},\ and\ \citenamefont
  {Keim}}]{Dillmann2012a}%
  \BibitemOpen
  \bibfield  {author} {\bibinfo {author} {\bibfnamefont {P.}~\bibnamefont
  {Dillmann}}, \bibinfo {author} {\bibfnamefont {G.}~\bibnamefont {Maret}},\
  and\ \bibinfo {author} {\bibfnamefont {P.}~\bibnamefont {Keim}},\ }\bibfield
  {title} {\bibinfo {title} {Comparison of 2d melting criteria in a colloidal
  system},\ }\href
  {https://doi.org/https://doi.org/10.1088/0953-8984/24/46/464118} {\bibfield
  {journal} {\bibinfo  {journal} {J. Phys. Cond. Mat.}\ }\textbf {\bibinfo
  {volume} {24}},\ \bibinfo {pages} {464118} (\bibinfo {year}
  {2012})}\BibitemShut {NoStop}%
\bibitem [{\citenamefont {Meel}\ \emph {et~al.}(2012)\citenamefont {Meel},
  \citenamefont {Filion}, \citenamefont {Valeriani},\ and\ \citenamefont
  {Frenkel}}]{Meel2012}%
  \BibitemOpen
  \bibfield  {author} {\bibinfo {author} {\bibfnamefont {J.~A.~v.}\
  \bibnamefont {Meel}}, \bibinfo {author} {\bibfnamefont {L.}~\bibnamefont
  {Filion}}, \bibinfo {author} {\bibfnamefont {C.}~\bibnamefont {Valeriani}},\
  and\ \bibinfo {author} {\bibfnamefont {D.}~\bibnamefont {Frenkel}},\
  }\bibfield  {title} {\bibinfo {title} {A parameter-free, solid-angle based,
  nearest-neighbor algorithm},\ }\href {https://doi.org/10.1063/1.4729313}
  {\bibfield  {journal} {\bibinfo  {journal} {The Journal of Chemical Physics}\
  }\textbf {\bibinfo {volume} {136}},\ \bibinfo {pages} {234107} (\bibinfo
  {year} {2012})}\BibitemShut {NoStop}%
\end{thebibliography}%

\end{document}